# Breaks in the Hall–Petch Relationship after Severe Plastic Deformation of Magnesium, Aluminum, Copper, and Iron

**Shivam Dangwal** [1,2], **Kaveh Edalati** [1,2,*], **Ruslan Z. Valiev** [3,4] **and Terence G. Langdon** [5]

[1] WPI International Institute for Carbon-Neutral Energy Research (WPI-I2CNER), Kyushu University, Fukuoka, Japan; dangwal.shivam.332@s.kyushu-u.ac.jp
[2] Department of Automotive Science, Graduate School of Integrated Frontier Sciences, Kyushu University, Fukuoka, Japan; kaveh.edalati@kyudai.jp
[3] Institute of Physics of Advanced Materials, Ufa University of Science and Technology, Ufa 450076, Russia; ruslan.valiev@ugatu.su
[4] Laboratory for Dynamics and Extreme Characteristics of Promising Nanostructured Materials, Saint Petersburg State University, Saint Petersburg 199034, Russia
[5] Materials Research Group, Department of Mechanical Engineering, University of Southampton, Southampton SO17 1BJ, U.K.; t.g.langdon@soton.ac.uk
* Correspondence: kaveh.edalati@kyudai.jp; Tel.: +81-82-802-6744

**Abstract:** Strengthening by grain refinement via the Hall–Petch mechanism and softening by nanograin formation via the inverse Hall–Petch mechanism have been the subject of argument for decades, particularly for ultrafine-grained materials. In this study, the Hall–Petch relationship is examined for ultrafine-grained magnesium, aluminum, copper, and iron produced by severe plastic deformation in the literature. Magnesium, aluminum, copper, and their alloys follow the Hall–Petch relationship with a low slope, but an up-break appears when the grain sizes are reduced below 500–1000 nm. This extra strengthening, which is mainly due to the enhanced contribution of dislocations, is followed by a down-break for grain sizes smaller than 70–150 nm due to the diminution of the dislocation contribution and an enhancement of thermally-activated phenomena. For pure iron with a lower dislocation mobility, the Hall–Petch breaks are not evident, but the strength at the nanometer grain size range is lower than the expected Hall–Petch trend in the submicrometer range. The strength of nanograined iron can be increased to the expected trend by stabilizing grain boundaries via impurity atoms. Detailed analyses of the data confirm that grain refinement to the nanometer level is not necessarily a solution to achieve extra strengthening, but other strategies such as microstructural stabilization by segregation or precipitation are required.

## 1. Introduction

Bulk nanostructured materials have ultrafine grains with sizes smaller than one micrometer with mainly high-angle grain boundaries, and they are receiving significant attention in recent years due to their enhanced mechanical and functional properties [1–3]. Severe plastic deformation (SPD) is currently the most effective technology for producing such bulk nanostructured materials [1–3]. Processing via SPD methods, such as high-pressure torsion (HPT) [4,5], equal-channel angular pressing (ECAP) [6,7], and accumulative roll-bonding (ARB) [8,9] not only refines grain sizes but also creates various kinds of lattice defects such as vacancies and dislocations [10,11]. An enhancement of mechanical properties such as strength and hardness is the most investigated feature of severely deformed materials, but there remain significant debates about the mechanisms of the strengthening of these materials [1–3]. While most studies consider that the high strength of severely deformed bulk nanostructured materials is due to the reduction of grain size [12–14], some studies suggest that the introduction of dislocations is the main strengthening mechanism in these materials [15,16]. The occurrence of thermally-activated phenomena and the formation of grain boundary segregation in nanostructured materials influence the strengthening mechanisms and make the evaluation of these materials even more complicated [17,18].

The strength of materials can be enhanced by several strategies such as solution hardening, dislocation hardening, grain-boundary hardening, and precipitation hardening [19]. For polycrystalline materials, the enhancement of yield stress ($\sigma_y$) by grain refinement can be expressed in terms of the average grain size ($d$) using the famous Hall–Petch equation [20,21].

$$\sigma_y = \sigma_0 + Ad^{-1/2} \qquad (1)$$

where $\sigma_0$ is the friction stress, and $A$ is a constant. The dislocation hardening can be quantified through the Bailey–Hirsch equation [22].

$$\sigma_y = M\alpha Gb\rho^{1/2} \qquad (2)$$

where $M$ is the Taylor factor of 3.06, $\alpha$ is a constant depending on the dislocation interaction, $G$ is the shear modulus, $b$ is the Burgers vector, and $\rho$ is the dislocation density. Some theoretical studies suggested that the dislocation density is inversely proportional to the grain size, and thus the variation of yield strength with grain size can follow a Hall–Petch-like relationship even in the presence of dislocations [19,23]. Therefore, it is expected that the slope of such a Hall–Petch-like relationship is influenced by the dislocation density [23,24]. The change of slope and appearance of an up-break in the Hall–Petch relationship was experimentally reported in some ultrafine-grained materials due to the contribution of dislocation hardening [25,26].

Another complication in the application of the Hall–Petch relationship to bulk nanostructured materials arises from the occurrence of the inverse Hall–Petch mechanism [27–29]. When the grain size is in the range of a few tens of nanometers, a down-break in the Hall–Petch relationship or a softening by the inverse Hall–Petch mechanism may occur due to the change of deformation mechanism from the dislocation activity to thermally-activated phenomena [30–32]. These thermally-activated phenomena, such as diffusional creep, grain-boundary sliding, or grain rotation, were experimentally observed at room temperature in some nanograined materials [33–35]. A large number of studies reported that some nanocrystalline materials follow down-breaks compared to the predictions of the Hall–Petch relationship [36–43]. Such down-breaks become more significant when the temperature is high or metals with low melting temperatures are examined [44]. In fact, softening with grain refinement was reported in severely deformed high-purity indium [45], tin [45], lead [45], zinc [46] and aluminum [47], and hardening by grain coarsening was reported in self-annealed magnesium [48]. Recent analyses by Figueiredo et al. suggested that room-temperature grain boundary sliding can contribute to the unusual softening behavior of these metals [49].

The occurrence of segregation due to non-equilibrium microstructural features of severely deformed materials is another parameter that can affect the strength of these materials [17,18]. Some studies attributed the large strength of severely deformed materials to the effect of segregation on the stability of grain boundaries and their interaction with dislocations [13,50]. It was shown that a combination of precipitation and segregation in aluminum alloys is quite effective to achieve high strengths close to 1 GPa [51]. It should be noted that the segregation does not necessarily lead to extra hardening, as it was shown that the segregation of some atoms can enhance thermally-activated phenomena, such as grain boundary sliding at room temperature, leading to softening rather than the expected hardening [52,53]. These studies suggest that, in addition to grain size, the nature of grain boundaries should be considered in the hardening/softening behavior of materials [54,55].

The grain sizes of severely deformed metals are usually in the range of the submicrometer level, but there are high interests to reduce the grain size further to the nanometer level (< 100 nm) to achieve enhanced mechanical properties through the Hall–Petch mechanism [1–3]. However, since grain boundaries can act as dislocation sinks in nanograined materials, they usually do not have high dislocation density, and this may negatively affect their strength [28]. Moreover, in addition to the cooperation of dislocation and grain boundary hardening, thermally-activated softening phenomena are likely to occur in these nanograined materials [33–35]. Therefore, it remains an open question whether the extension of grain refinement to the nanometer level using SPD processing is favorable for achieving ultrahigh strength and hardness. To shed light on this open question, an examination of the Hall–Petch relationship from the micrometer grain sizes to the submicrometer grain sizes through the nanograin sizes is beneficial.

In this study, the Hall–Petch relationship for magnesium, aluminum, copper, and iron after processing with SPD was investigated. It was found that the grain refinement to the nanometer level does not necessarily lead to higher strength and hardness due to the breaks in the Hall–Petch relationship, but other strategies such as stabilization of grains and grain boundaries by precipitates or segregation are required.

## 2. Experimental Data

This study investigates the Hall–Petch relationship for four groups of materials: (i) magnesium and its alloys, which have low melting points and hexagonal close-packed (HCP) structure; (ii) aluminum and its alloys, which have low melting points and face-centered cubic (FCC) structure; (iii) copper and its alloys, which have moderate melting points and an FCC crystal structure; and (iv) iron with a high melting point and a body-centered cubic (BCC) structure. By considering the melting points and crystal structures, the fastest dislocation mobility belongs to magnesium and aluminum, and iron has the slowest dislocation mobility. The experimental data for yield strength, hardness and grain size were gathered from the literature. The grain sizes (the average diameter of the regions separated by high-angle grain boundaries) were measured either with transmission electron microscopy (TEM) or electron back-scatter diffraction (EBSD). To plot the Hall–Petch relationship, the hardness values were converted to the yield stress by the Tabor equation [56].

$$\sigma_y \text{ (MPa)} = HV \text{ (MPa)} / 3 = 9.81 \times HV \text{ (Hv)} / 3. \qquad (3)$$

The gathered data are either after HPT, ECAP, or ARB processes, but some data for the annealed and coarse-grained materials were included for comparison. For magnesium, since texture can influence the data after ECAP and ARB processing, only the data after HPT processing were gathered. Moreover, since iron alloys can show phase transformations or the presence of other compounds such as carbides, they were not included in this study to avoid complications. The composite materials or materials containing particles were also excluded, but limited age-hardenable aluminum alloys were included for comparison. The gathered data, taken from Refs. [57–157], are given in Tables A1, A2, A3, and A4 of the Appendix A for magnesium, aluminum, copper, and iron, respectively.

## 3. Results

### 3.1. Magnesium and Its Alloys

The plot of yield stress versus the inverse square root of grain size for magnesium and its alloys is shown in Figure 1. The datum points for coarse-grained pure magnesium follow the Hall–Petch relation with a low slope until the grain size reaches ~1000 nm. For the grain sizes smaller than 1000 nm, which are achieved for only SPD-processed magnesium alloys, there is an up-break in the Hall–Petch relationship, and the strength increases faster in this region. The scattering of datum points in the range of 100-1000 nm is quite significant, suggesting the non-uniform contribution of other strengthening mechanisms such as dislocation hardening. Such scattering can also be partly due to inconsistencies in the experimental grain size measurement methods as well as due to a lack of precision in the measurement methods. When grain sizes are reduced further, a down-break appears roughly at ~150 nm, and the reduction of the grain size below 100 nm does not significantly change the strength. It should be noted that the drawn lines are a rough fitting of the available data, but much more data are required to determine the exact grain sizes in which the breaks of the Hall–Petch relationship occur.

### 3.2. Aluminum and Its Alloys

For aluminum and its alloys, the trend of datum points in the Hall–Petch plot is rather similar to magnesium, although the scattering of the data is less, as shown in Figure 2. The strength increases with decreasing grain size in the micrometer grain size ranges, but the slope of the plot increases when the grain sizes are at the submicrometer level. This up-break is followed by a down-break when the grain sizes are changed from the submicrometer level to the nanometer level. The highest strength values are achieved for aluminum alloys with submicrometer grain sizes containing precipitates. Moreover, the presence of segregation in some nanograined alloys, such as Al-Ca, leads to a positive deviation of strength in the nanometer grain size region. Taken altogether, Figure 2 confirms that the Hall–Petch relationship for aluminum and its alloys is quite different for grain sizes at the micrometer, submicrometer, and nanometer levels, suggesting that different deformation and hardening mechanisms become dominant in aluminum depending on the grain size.

*3.3. Copper and Its Alloys*

The plot between the strength and the inverse square root of the grain size for copper and its alloys is shown in Figure 3. The datum points scatter significantly, and it is hard to establish a clear Hall–Petch relationship. However, one can conclude that the datum points for coarse-grained copper follow a Hall–Petch relationship with a slope like the one suggested by Smith and Hashemi for copper ($k$ = 110 MPa. $\mu m^{1/2}$) [158]. For grain sizes larger than 500 nm, there are deviations in the datum points to higher strength values, and the maximum deviations are achieved for a grain size of 70 nm. With a further reduction in the grain size below 70 nm, the negative deviations appear, and the increase in the strength becomes less significant.

*3.4. Iron*

The Hall–Petch plot for iron is shown in Figure 4. Despite some deviations, the datum points fall close to the Hall–Petch relationship suggested by Takaki et al. for pure iron with $\sigma_0$ = 100 MPa and $k$ = 600 MPa. $\mu m^{1/2}$ [159,160]. The largest deviation from this plot is achieved for a ball-milled pure iron sample with grain sizes smaller than 30 nm. However, even for these small grain sizes, the deviation from the plot becomes small by the stabilization of grain boundaries using the carbon atom segregation, as attempted by Borchers et al. [155]. In conclusion, compared to magnesium, aluminum, and copper, no clear evidence for the breaks of the Hall–Petch relationship is found for pure iron, which has a relatively high melting temperature and slow dislocation mobility.

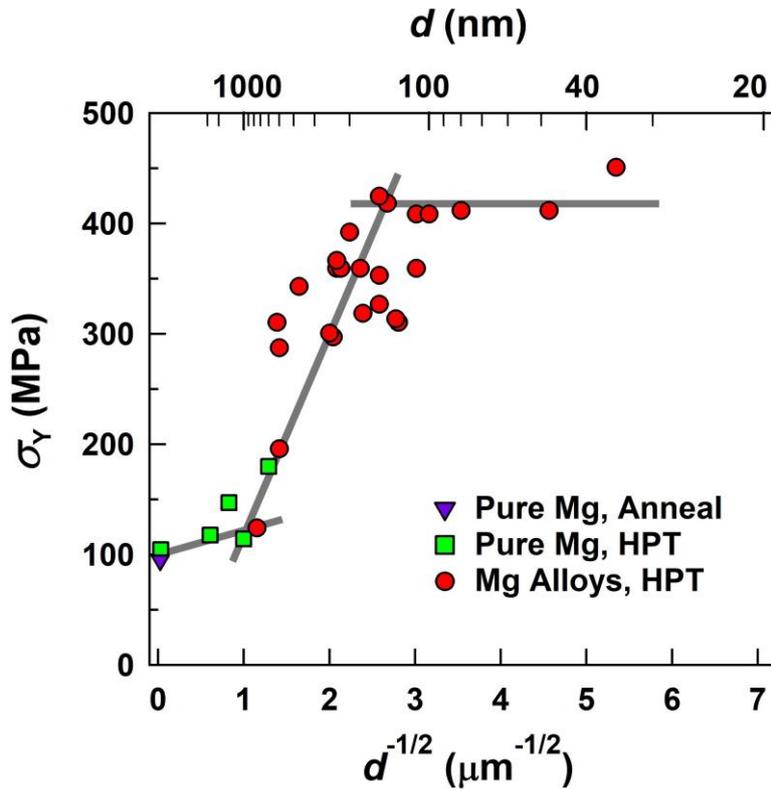

**Figure 1.** The Hall–Petch plot (yield stress, $\sigma_y$, versus inverse square root of grain size, $d$) and its breaks for magnesium and its alloys before and after severe plastic deformation by high-pressure torsion (HPT).

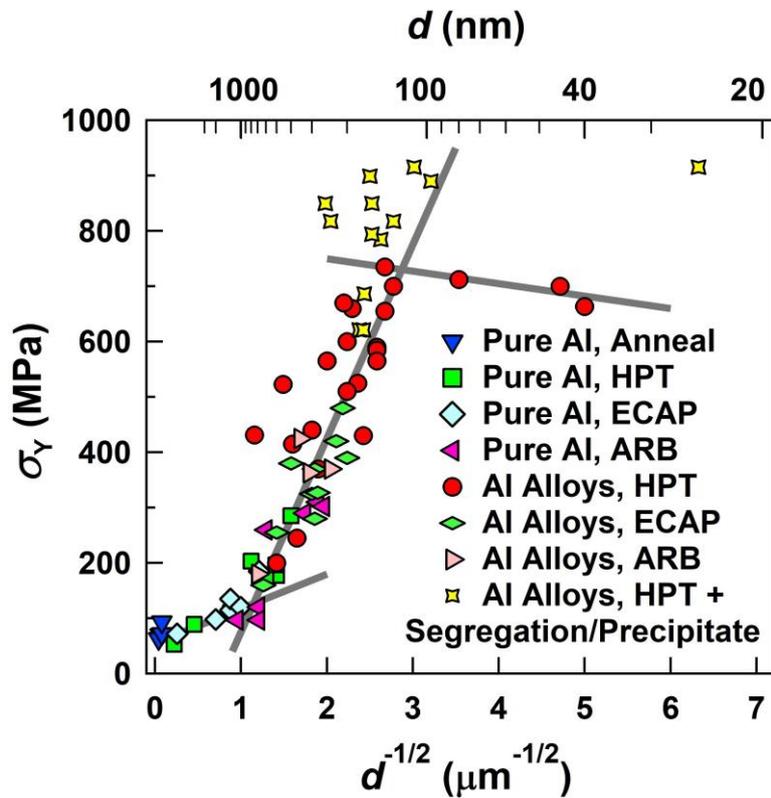

**Figure 2.** The Hall–Petch plot (yield stress, $\sigma_y$, versus inverse square root of grain size, $d$) and its breaks for aluminum and its alloys before and after severe plastic deformation by high-pressure torsion (HPT), equal-channel angular pressing (ECAP), and accumulative roll-bonding (ARB).

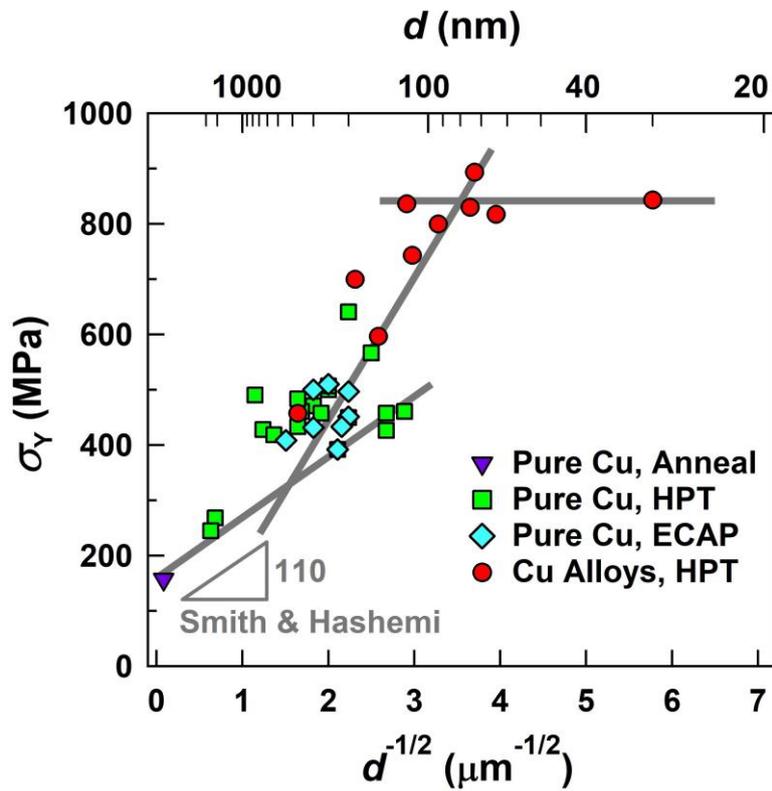

**Figure 3.** The Hall–Petch plot (yield stress, $\sigma_y$, versus inverse square root of grain size, $d$) and its breaks for copper and its alloys before and after severe plastic deformation by high-pressure torsion (HPT) and equal-channel angular pressing (ECAP).

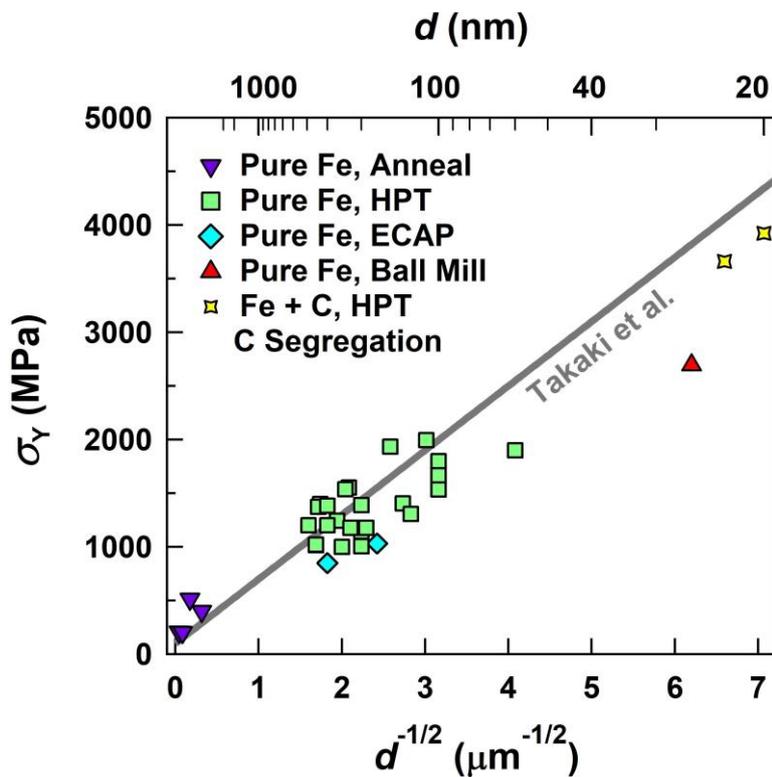

**Figure 4.** The Hall–Petch plot (yield stress, $\sigma_y$, versus inverse square root of grain size, $d$) for iron before and after severe plastic deformation by high-pressure torsion (HPT) and equal-channel angular pressing (ECAP) in comparison with the Hall–Petch relation suggested by Takaki et al. [159,160].

## 4. Discussion

The current study shows different behaviors of magnesium, aluminum, copper, and iron when their yield stress is examined in different grain size regions through the Hall–Petch plots. While magnesium, aluminum, and copper show one up-break and one down-break in the Hall–Petch relationship, iron fails to show a clear break. Three questions need to be discussed. (i) What is the reason for the occurrence of such breaks? (ii) What is the reason for the different behaviors of the four selected metallic groups? (iii) Is grain refinement to the nanometer level beneficial to achieve ultrahigh strength in severely deformed materials?

Regarding the first question, it should be noted that the strength of materials is influenced by different mechanisms such as solution hardening, dislocation hardening, grain-boundary hardening, and precipitation hardening [19]. An earlier study suggested that the effect of solution hardening compared to other hardening mechanisms is smaller and can be ignored in severely deformed metals and single-phase alloys [93]. The study suggested that the most significant effect of solute atoms was the grain size reduction by enhancing the dislocation–solute atom interactions. Precipitation hardening has a significant effect on the strength [103–107], but most data gathered for this study were in the absence of any precipitates or second-phase particles (except for some age-hardenable aluminum alloys in Figure 2). Therefore, the two main strengthening mechanisms are expected to influence the trend of the data in Figures 1-4: grain boundary hardening and dislocation hardening. For coarse-grained materials, and particularly for the annealed ones, the dislocation density is not significant and the strength can be explained by a classic Hall–Petch relationship with a low slope [20,21]. When grain sizes are reduced below some levels (usually to the submicrometer range) by SPD processing, large numbers of dislocations are also introduced [11,15,16] and extra hardening and an up-break in the Hall–Petch relationship are observed. With a further reduction of grain size below some critical levels, on the one hand, the dislocation density is usually reduced because nanograins can act as dislocation sinks [28]. On the other hand, thermally-activated phenomena can be activated in nanograined materials even at room temperature, particularly in aluminum and magnesium [27–43]. As mentioned earlier, softening by grain refinement and hardening by grain coarsening were reported in high-purity HPT-processed magnesium (99.9%) and aluminum (99.9999%) as an indication of thermally-activated phenomena [47,48]. Moreover, creep analyses conducted by Figueiredo et al. [49] suggest that grain boundary sliding can play a significant role in the deformation of low-melting-temperature metals. Moreover, grain-boundary sliding was experimentally observed in severely deformed metals, including in pure copper [34]. Therefore, as a result of dislocation annihilation and the operation of thermally-activated phenomena, a down-break appears in the Hall–Petch relationship for magnesium, aluminum, and copper in Figures 1-3. It is concluded that the examination of the Hall–Petch relationship for severely deformed materials, without considering the contribution of dislocations, does not readily reflect the grain boundary strengthening even if the data reasonably follow a Hall–Petch-like relationship.

Regarding the second question, iron apparently follows the Hall–Petch relationship, but the three other metals show breaks in the Hall–Petch relationship. Here, it should be noted that only pure iron was included in this study, because iron alloys are rather complicated in their structure and strengthening mechanisms. For iron, which has the strongest atomic bond energy, the highest melting point, and the slowest dislocation mobility among the four selected metallic systems [161], there should be a reasonable inverse relation between the dislocation density and the grain size, leading to a Hall–Petch-like relationship from the micrometer grain sizes to the nanometer grain sizes [23]. However, for magnesium and aluminum with low melting temperatures and for copper with a moderate melting temperature, the dislocation mobility is faster, and it is influenced more significantly by grain size and solute atoms [93]. The down-break in the Hall–Petch relationship, which is due to the annihilation of dislocations in the nanograin boundaries [28] and/or due to the possible contribution of thermally-activated phenomena [33–35], depends on the mobility of the dislocations in the alloys. The down-break appears at about ~150 nm for magnesium with the highest dislocation mobility, ~100 nm for aluminum, and ~70 nm for copper in Figures 1–3. For iron with the slowest mobility of dislocations, although the largest negative deviation from the Hall–Petch relationship occurs at grain sizes below 30 nm in Figure 4, its Hall–Petch behavior at the nanometer level needs to be studied in the future by producing nanograined samples. Here, it should be mentioned again that the lines in Figures 1–4 are rough fitting of the available data, and more data should be gathered in the future to determine grain sizes in which the Hall–Petch relationship exhibits up-breaks and down-breaks.

Regarding the third question on the significance of grain refinement to the nanometer region on extra strengthening, each material can be discussed separately. For magnesium, the reduction of grain size to the nanometer level does not provide a clear benefit to achieve extra strengthening. This negative behavior can be overcome by adding a large concentration of solute atoms into magnesium to suppress the dislocation mobility, as attempted earlier by using the concept of ultra-SPD [162,163]. It was shown that highly alloyed magnesium alloys with 10 nm grain size can have hardness levels of 360 Hv, corresponding to an ultrahigh yield strength of 1080 MPa [162,163]. For aluminum, the maximum hardness is achieved for the submicrometer grain sizes, particularly in the presence of precipitates and/or segregation [13,50,51]. At the nanometer level, the hardness of aluminum can still be enhanced if grain boundaries are stabilized by segregation, as attempted by Sauvage et al. for the nanograined Al-Ca alloy [97]. For copper, a reduction of the grain size to 70 nm seems quite effective to achieve high strength, but a further reduction in the grain size does not necessarily lead to extra strengthening. For iron, the down-break in the Hall–Petch relationship may occur for small nanograin sizes such as 30 nm, but, even in nanograined iron, stabilization of grain boundaries by segregation is effective to achieve extra hardening, as attempted by Borchers et al. [155]. Taken altogether, to have extra hardening in nanocrystalline materials, grain refinement should be accompanied by other strategies such as the modification of the nature of grain boundaries and the stabilization of the microstructure by segregation or precipitates [54,55,164]. Thus, the grain size alone is not the determining factor for strengthening, but the features of grain boundaries and grains should be considered to achieve ultrahigh strength.

## 5. Conclusions

The validity of the Hall–Petch relationship for severely deformed magnesium, aluminum, copper, and their alloys, as well as for iron, was examined. Up-breaks due to the enhanced contribution of dislocation hardening and down-breaks due to the reduced contribution of dislocation hardening and the possible contribution of thermally-activated phenomena were observed in magnesium, aluminum, and copper. Iron mainly followed the Hall–Petch relationship from the micrometer to nanometer grain size ranges, although the hardening at the nanometer grain sizes somehow negatively deviated from the relationship. To achieve a high strength, a combination of grain refinement and precipitation/segregation generation appears to be more effective than only nanograin formation. Future studies are required to quantitatively analyze the contribution of different hardening mechanisms to the strength of severely deformed nanograined materials to clarify the role of the inverse Hall–Petch relationship in their mechanical behavior.

## Appendix A

Data taken from the literature for grain size (measured either with TEM or EBSD), yield stress or hardness divided by three, and dislocation density are given in Table 1 for magnesium and its alloys, Table 2 for aluminum and its alloys, Table 3 for copper and its alloys, and Table 4 for iron before and after processing using SPD methods.

**Table 1.** Reported experimental data for grain size, yield strength or hardness divided by three, and dislocation density for magnesium and its alloys before and after processing by severe plastic deformation.

| Material | Process | Grain Size (μm) | $HV/3$ or $\sigma_y$ (MPa) | Dislocation Density (m$^{-2}$) | Ref. |
|---|---|---|---|---|---|
| Mg (99.9%) | Anneal | 1600 | 94 | | [46] |
| Mg (99.8%) | As-Cast | 1000 | 104 | | [57] |
| Mg (99.9%) | HPT | 2.7 | 117 | | [46] |
| Mg (99.9%) | HPT | 1 | 114 | | [58] |

| Material | Process | Grain Size (μm) | $HV/3$ or $\sigma_y$ (MPa) | Dislocation Density (m$^{-2}$) | Ref. |
|---|---|---|---|---|---|
| Mg (99.8%) | HPT | 0.6 | 179 | | [57] |
| Pure Mg | HPT | 1.46 | 147 | | [59] |
| Mg-8Li (wt%) | HPT | 0.5 | 196 | | [60] |
| Mg-9Al (wt%) | HPT | 0.15 | 353 | | [61] |
| Mg-0.41Dy (wt%) | HPT | 0.75 | 124 | 3.5×10$^{14}$ | [62] |
| Mg-3.4Zn (at%) | HPT | 0.14 | 418 | | [63] |
| Mg-4.3Zn-0.7Y (at%) | HPT | 0.15 | 425 | 5×10$^{14}$ | [64] |
| Mg-1Zn-0.13Ca (wt%) | HPT | 0.15 | 327 | | [65] |
| Mg-8Gd-3Y-0.4Zr (wt%) | HPT | 0.08 | 412 | | [66] |
| Mg-8Gd-3.8Y-1Zn-0.4Zr (wt%) | HPT | 0.048 | 412 | 4.65×10$^{14}$ | [67] |
| AM60 | HPT | 0.23 | 366 | 11×10$^{14}$ | [68] |
| AZ31 | HPT | 0.5 | 287 | | [69] |
| AZ31 | HPT | 0.11 | 408 | | [70] |
| AZ31 | HPT | 0.175 | 318 | 5.7×10$^{14}$ | [71] |
| AZ61 | HPT | 0.52 | 310 | | [72] |
| AZ61 | HPT | 0.37 | 343 | | [72] |
| AZ61 | HPT | 0.23 | 359 | | [72] |
| AZ61 | HPT | 0.22 | 359 | | [72] |
| AZ61 | HPT | 0.11 | 359 | | [72] |
| AZ80 | HPT | 0.2 | 392 | | [73] |
| AZ80 | HPT | 0.1 | 408 | | [74] |
| AZ91 | HPT | 0.18 | 359 | 1.1×10$^{14}$ | [75] |
| AZ91 | HPT | 0.25 | 300 | 0.6×10$^{14}$ | [75] |
| AZ91 | HPT | 0.035 | 451 | 5×10$^{14}$ | [75] |
| EZ33A | HPT | 0.24 | 297 | | [76] |
| EZ33A | HPT | 0.127 | 310 | | [76] |
| EZ33A | HPT | 0.13 | 313 | | [76] |
| ZKX600 | HPT | 0.1 | 408 | | [77] |

**Table 2.** Reported experimental data for grain size, yield strength or hardness divided by three, and dislocation density for aluminum and its alloys before and after processing by severe plastic deformation.

| Material | Process | Grain Size (μm) | $HV/3$ or $\sigma_y$ (MPa) | Dislocation Density (m$^{-2}$) | Ref. |
|---|---|---|---|---|---|
| Al (99.9999%) | Anneal | >1000 | 60.5 | | [47] |
| Al (99.999%) | Anneal | >1000 | 61.1 | | [47] |
| Al (99.99%) | Anneal | 620 | 63.1 | | [47] |
| Al (99.7%) | Anneal | 300 | 69.9 | | [78] |
| Al (99.5%) | Anneal | 170 | 72.0 | | [47] |
| Al (99%) | Anneal | 160 | 92.9 | | [47] |
| Al (99.9999%) | HPT | 20 | 53 | | [47] |
| Al (99.999%) | HPT | 4.8 | 89 | | [47] |
| Al (99.99%) | HPT | 1.3 | 110 | | [47] |
| Al 99.7% | HPT | 0.8 | 203.2 | | [78] |
| Al (99.5%) | HPT | 0.5 | 177 | 2-4×10$^{14}$ | [79] |
| Al (99.5%) | HPT | 0.6 | 173 | | [47] |
| Al (99.5%) | HPT | 0.504 | 197 | 0.69×10$^{13}$ | [80] |
| Al (99%) | HPT | 0.4 | 285 | | [47] |
| Al (99.999%) | ECAP | 15 | 72 | | [81] |

| Material | Process | Grain size (μm) | Strength (MPa) | Dislocation density (m$^{-2}$) | Ref. |
|---|---|---|---|---|---|
| Al (99.999%) | ECAP | 1.3 | 115 | | [82] |
| Al (99.99%) | ECAP | 1.3 | 135 | | [83] |
| Al (99.99%) | ECAP | 2 | 98 | | [84] |
| Al (99.99%) | ECAP | 1 | 120 | 1.8×10$^{14}$ | [85] |
| Al (99%) | ECAP | 0.68 | 185 | | [86] |
| Al (99.99%) | ARB | 1.1 | 97 | 0.12×10$^{14}$ | [87] |
| Al (99.99%) | ARB | 0.69 | 120 | 0.12×10$^{14}$ | [88] |
| Al (99.99%) | ARB | 0.69 | 97 | 0.123×10$^{14}$ | [26] |
| A1 (99.5%) | ARB | 0.6 | 260 | 1.33×10$^{14}$ | [89] |
| Al (99%) | ARB | 0.33 | 290 | | [90] |
| Al (99%) | ARB | 0.28 | 310 | 1.33x10$^{14}$ | [88] |
| Al (99%) | ARB | 0.26 | 302 | | [91] |
| Al-1.1Mg (at%) | HPT | 0.39 | 415 | | [92,93] |
| Al-3.3Mg (at%) | HPT | 0.2 | 600 | | [92,93] |
| Al-5.5Mg (at%) | HPT | 0.19 | 660 | | [92,93] |
| Al-8.8Mg (at%) | HPT | 0.14 | 735 | | [92,93] |
| Al-3Mg-0.2Sc (wt%) | HPT | 0.15 | 585 | | [94] |
| Al-3Mg (wt%) | ECAP | 0.2 | 390 | 27×10$^{14}$ | [85] |
| Al-3Mg-0.2Sc (wt%) | HPT | 0.15 | 565.3 | | [95] |
| Al-5.9Mg-0.3Sc-0.18Zr (wt%) | HPT | 0.045 | 700 | | [96] |
| Al-Ca | HPT | 0.025 | 915.32 | | [97] |
| Al-1.7Cu (at%) | HPT | 0.207 | 670 | | [92,93] |
| Al-3Cu-1Li (wt%) | HPT | 0.13 | 700 | 27.8×10$^{14}$ | [98] |
| Al-2Fe (wt%) | HPT | 0.3 | 440 | <15×10$^{14}$ | [99] |
| Al-2Fe (wt%) | HPT | 0.15 | 590 | <26×10$^{14}$ | [99] |
| Al-2Fe (wt%) | HPT | 0.14 | 655 | <33×10$^{14}$ | [99] |
| Al-1.3Ag (at%) | HPT | 0.5 | 200 | | [92,93] |
| Al-3.0Ag (at%) | HPT | 0.367 | 245 | | [92,93] |
| Al-5.9Ag (at%) | HPT | 0.278 | 370 | | [92,93] |
| Al-5Zr (wt%) | HPT | 0.073 | 915.32 | >1×10$^{14}$ | [100] |
| Al-3.1La-5.4Ce | HPT | 0.040 | 663.3 | 1.3×10$^{14}$ | [101] |
| Al-3.1La-5.4Ce | HPT | 0.08 | 712.35 | | [101] |
| A1570 | HPT | 0.097 | 890 | | [13] |
| A2024 | ECAP | 0.3 | 325 | | [86] |
| A2024 | HPT | 0.13 | 817.25 | 55×10$^{14}$ | [102] |
| A2024 | HPT | 0.16 | 898.975 | 10×10$^{14}$ | [102] |
| A2024 | HPT | 0.24 | 817.25 | 8×10$^{14}$ | [103] |
| A2024 | HPT | 0.255 | 849.94 | 5×10$^{14}$ | [103] |
| A2024 | HPT | 0.24 | 817.25 | | [104] |
| A2024 | HPT | 0.177 | 621.11 | 2.7x10$^{14}$ | [104] |
| A2024 | HPT | 0.145 | 784.56 | 3.3×10$^{14}$ | [104] |
| A2024 | HPT | 0.157 | 849.94 | | [104] |
| AA2024 | HPT | 0.157 | 490.35 | 3.2×10$^{14}$ | [105] |

| Material | Process | | | | Ref. |
|---|---|---|---|---|---|
| AA2024 | HPT | 0.169 | 686.49 | $2.2\times10^{14}$ | [105] |
| A6060 | HPT | 0.18 | 525 | | [106] |
| A6061 | HPT | 0.2 | 510 | $2.6\times10^{14}$ | [107] |
| A6061 | HPT | 0.17 | 430 | | [108] |
| Al6061 | HPT | 0.45 | 522.8 | | [109] |
| Al6061 | HPT | 0.746 | 431.3 | | [109] |
| Al6061 | HPT | 0.25 | 565.3 | | [110] |
| AA7075 | HPT | 0.11 | 915.32 | | [50] |
| AA7075 | HPT | 0.17 | 621.1 | | [50] |
| Al-0.2Zr (wt%) | ECAP | 0.63 | 160 | | [111] |
| A3004 | ECAP | 0.29 | 370 | | [86] |
| A5083 | ECAP | 0.225 | 420 | | [86] |
| A6061 | ECAP | 0.4 | 380 | | [112] |
| A6061 | ECAP | 0.29 | 280 | | [86] |
| A6061 | ECAP | 0.28 | 327 | | [83] |
| A6063 | ECAP | 0.5 | 255 | | [106] |
| A7075 | ECAP | 0.21 | 480 | | [86] |
| A2024 | ARB | 0.35 | 425 | | [113] |
| A6061 | ARB | 0.24 | 370 | | [114] |
| A6061 | ARB | 0.31 | 363 | | [115] |
| A8011 | ARB | 0.7 | 180 | | [116] |

**Table 3.** Reported experimental data for grain size, yield strength or hardness divided by three, and dislocation density for copper and its alloys before and after processing by severe plastic deformation.

| Material | Process | Grain Size (μm) | $HV/3$ or $\sigma_y$ (MPa) | Dislocation Density (m$^{-2}$) | Ref. |
|---|---|---|---|---|---|
| Cu (99.99%) | Anneal | 150 | 156 | | [117] |
| Cu (99.99%) | HPT+Anneal | 2.15 | 268 | | [117] |
| Cu (99.99%) | HPT+Anneal | 2.5 | 245 | | [117] |
| Cu (99.99%) | HPT | 0.37 | 433 | | [92,93] |
| Cu (99.99%) | HPT | 0.25 | 500 | 43.4×10$^{14}$ | [118] |
| Cu (99.99%) | HPT | 0.3 | 470 | | [119] |
| Cu (99.99%) | HPT | 0.225 | 392 | | [120] |
| Cu (99.99%) | HPT | 0.273 | 474 | 0.39×10$^{14}$ | [80] |
| Cu (99.98%) | HPT | 0.2 | 449 | | [121] |
| Cu (99.98%) | HPT | 0.16 | 566 | 37×10$^{14}$ | [122] |
| Cu (99.97%) | HPT | 0.14 | 457 | | [123] |
| Cu (99.97%) | HPT | 0.12 | 461 | | [124] |
| Cu (99.96%) | HPT | 0.14 | 426 | | [125] |
| Cu (99.95%) | HPT | 0.25 | 506 | 70×10$^{14}$ | [126] |
| Cu (99.9%) | HPT | 0.76 | 490 | | [127] |
| Cu (99.9%) | HPT | 0.2 | 640 | | [119] |
| Cu (99.9%) | HPT | 0.65 | 428 | | [128] |
| Cu (99.9%) | HPT | 0.35 | 461 | | [128] |
| Cu (99.87%) | HPT | 0.537 | 418 | 1.48×10$^{14}$ | [129] |
| Cu (99.87%) | HPT | 0.37 | 483 | 8.6×10$^{14}$ | [130] |
| Cu (99.98%) | ECAP+HPT | 0.225 | 392 | | [121] |
| Cu (99.99%) | ECAP | 0.3 | 431 | | [117] |
| Cu (99.98%) | ECAP | 0.215 | 433 | | [122] |
| Cu (99.98%) | ECAP | 0.25 | 510 | 4.09×10$^{14}$ | [131] |
| Cu (99.95%) | ECAP | 0.44 | 408 | | [132] |
| Cu (99.95%) | ECAP | 0.2 | 497 | | [133] |
| Cu (99.9%) | ECAP | 0.2 | 451 | | [134] |
| Cu (99.9%) | ECAP | 0.3 | 500 | | [119] |
| Cu-4.6Al (at%) | HPT | 0.187 | 700 | | [92,93] |
| Cu-11Al (at%) | HPT | 0.118 | 836 | | [92,93] |
| Cu-15Al (at%) | HPT | 0.073 | 893 | | [92,93] |
| Cu-16Al (at%) | HPT | 0.03 | 843 | | [124] |
| Cu-1.49Si (wt%) | HPT | 0.15 | 597 | 5×10$^{14}$ | [135] |
| Cu-9.7Zn (at%) | HPT | 0.113 | 743 | | [92,93] |
| Cu-19.5Zn (at%) | HPT | 0.093 | 800 | | [92,93] |
| Cu-29.4Zn (at%) | HPT | 0.075 | 830 | | [92,93] |
| Cu-30Zn (wt%) | HPT | 0.064 | 810 | 5.67×10$^{14}$ | [80] |
| Cu-0.17Zr (wt%) | ECAP | 0.37 | 457 | | [1376 |

**Table 4.** Reported experimental data for grain size, yield strength or hardness divided by three, and dislocation density for iron before and after processing by severe plastic deformation.

| Material | Process | Grain Size (μm) | $HV/3$ or $\sigma_y$ (MPa) | Dislocation Density (m$^{-2}$) | Ref. |
|---|---|---|---|---|---|
| Fe (99.95%) | Anneal | 500 | 205 | | [137] |
| Fe (99.96%) | None | 140 | 202 | | [138] |
| Fe (99.94%) | None | 33 | 513 | | [138] |
| Fe (99.88%) | None | 10 | 398 | | [138] |
| Fe (99.9988%) | HPT | 0.35 | 1016 | | [139] |
| Fe (99.99%) | HPT | 0.2 | 1144 | | [140] |
| Fe (99.99%) | HPT | 0.265 | 1242 | | [141] |
| Fe (99.97%) | HPT | 0.06 | 1900 | | [142] |
| Fe (99.96%) | HPT | 0.35 | 1020 | | [138] |
| Fe (99.96%) | HPT | 0.226 | 1184 | $0.97 \times 10^{14}$ | [80] |
| Fe (99.96%) | HPT | 0.2 | 1006 | | [14] |
| Fe (99.95%) | HPT | 0.195 | 1177 | | [143] |
| Fe (99.95%) | HPT | 0.19 | 1170 | | [137] |
| Fe (99.95%) | HPT | 0.3 | 1200 | | [144] |
| Fe (99.94%) | HPT | 0.34 | 1373 | | [138] |
| Fe (99.88%) | HPT | 0.23 | 1553 | | [138] |
| Fe (99.88%) | HPT | 0.24 | 1536 | | [138] |
| Fe (97.78%) | HPT | 0.11 | 1994 | | [138] |
| Pure Iron | HPT | 0.39 | 1200 | | [145] |
| Pure Iron | HPT | 0.15 | 1933 | | [146] |
| Armco Iron | HPT | 0.1 | 1667 | | [142] |
| Armco Iron | HPT | 0.134 | 1406 | | [147] |
| Armco Iron | HPT | 0.1 | 1533 | | [148] |
| Armco Iron | HPT | 0.25 | 1000 | | [149] |
| Armco Iron | HPT | 0.3 | 1383 | | [150] |
| Armco Iron | HPT | 0.2 | 1389 | | [151] |
| Fe-0.01C (wt%) | HPT | 0.125 | 1308 | $22 \times 10^{14}$ | [152] |
| Fe-0.02C (wt%) | HPT | 0.33 | 1400 | | [153] |
| Fe-0.03C (wt%) | HPT | 0.1 | 1800 | | [154] |
| Fe C | Milling + HPT | 0.023 | 3662 | | [155] |
| Fe C | Milling + HPT | 0.02 | 3924 | | [155] |
| Fe (99.95%) | ECAP | 0.3 | 850 | | [156] |
| Armco Iron | ECAP | 0.17 | 1031 | | [157] |
| Fe (97.78%) | Milling | 0.026 | 2697 | | [138] |

**Author Contributions:** Conceptualization, S.D., K.E., R.Z.V. and T.G.L.; writing—review and editing, S.D., K.E., R.Z.V. and T.G.L. All authors have read and agreed to the published version of the manuscript.

**Funding:** This work was supported, in part, by the MEXT, Japan, through Grants-in-Aid for Scientific Research (JP19H05176, JP21H00150, and JP22K18737). R.Z.V. gratefully acknowledges the financial support from the Russian Science Foundation in the framework of the Project No. 22-19- 00445. The work of TGL was supported by the European Research Council under ERC Grant Agreement No. 267464-SPDMETALS.

**Institutional Review Board Statement:** Not applicable.

**Informed Consent Statement:** Not applicable.

**Data Availability Statement:** No new data were created or analyzed in this study. Data sharing is not applicable to this article.

**Conflicts of Interest:** The authors declare no conflicts of interest.